\documentclass[doublecol]{arXiv} 

\usepackage{dcolumn}    
\usepackage{bm} 
\usepackage{graphicx}
\usepackage{amsmath}    
\usepackage{latexsym}
\usepackage{amsfonts}   
\usepackage{amssymb}
\usepackage{array}      
\usepackage{epsfig}
\usepackage[colorlinks=true,linkcolor=blue,urlcolor=blue,citecolor=blue]{hyperref}
\usepackage{color}
\usepackage{hyperref}
\usepackage[normalem]{ulem}

\newcommand{\bla}{bla\\bla\\bla\\bla\\bla}

\begin{document}
\title{Collision models in open system dynamics:\\A versatile tool for deeper insights?}

\author{Steve Campbell$^{1,2}$ and Bassano Vacchini$^{3,4}$}

\institute{$^1$School of Physics, University College Dublin, Belfield, Dublin 4, Ireland \\
$^2$Centre for Quantum Engineering, Science, and Technology, University College Dublin, Belfield, Dublin 4, Ireland \\
$^3$Dipartimento di Fisica ``Aldo Pontremoli”, Universit\`a degli Studi di Milano, via Celoria 16, 20133 Milan, Italy \\
$^4$Istituto Nazionale di Fisica Nucleare, Sezione di Milano, via Celoria 16, 20133 Milan, Italy}

\abstract{Understanding and simulating how a quantum system interacts and exchanges information or energy with its surroundings is a ubiquitous problem, one which must be carefully addressed in order to establish a coherent framework to describe the dynamics and thermodynamics of quantum systems. Significant effort has been invested in developing various methods for tackling this issue and in this Perspective we focus on one such technique, namely collision models, which have emerged as a remarkably flexible approach. We discuss their application to understanding non-Markovian dynamics and to studying the  thermodynamics of quantum systems, two areas in which collision models have proven to be particularly insightful. Their simple structure endows them with extremely broad applicability which has spurred their recent experimental demonstrations. By focusing on these areas, our aim is to provide a succinct entry point to this remarkable framework.}
\date{\today}
\maketitle

\section{Introduction} 
Try as we might, we can never fully prevent a quantum system from interacting with its surroundings. This simple observation motivates the need to develop a framework for assessing how a given system exchanges information and/or energy with other degrees of freedom i.e. its environment. Drastic simplifications can be achieved if, for example, the system and environment are sufficiently distinct in terms of evolution time-scales and relative size. Under such circumstances the environment acts as a sink into which information and energy can flow irretrievably. This process characterises as a memoryless or ``Markovian" dynamics and a closed dynamical equation of motion for the system can be expressed in the celebrated GKSL form~\cite{BreuerBook}. While this represents an extremely powerful approach, it nevertheless leaves significant questions unanswered. Indeed in general, one cannot assume that the system and environment remain uncorrelated throughout the dynamics or that the system has no effect on the environment's evolution. Thus, there has been a significant effort invested into developing a more complete description of open quantum system dynamics that capture non-Markovian features~\cite{Rivas2014a,Breuer2016a}.

Beyond studying the dynamics of a quantum system, understanding exchanges of information and energy between a system and its environment is also crucial in order to devise a proper thermodynamic framework. While quantum definitions of work and heat have been developed, these concepts are still being refined for generic dynamics. For simple Markovian settings, a meaningful thermodynamic description can be established by focusing solely on the system's degrees of freedom~\cite{Alicki}. However, this is no longer true for more general dynamics, e.g. strong coupling and non-Markovian cases, where the environment plays a more active role in the ensuing evolution.

It should be clear that for an arbitrary setting, the assessment of the dynamics and the thermodynamics necessitates at least partial information regarding the environmental degrees of freedom. In this regard, collision models (CMs), also equivalently known as repeated interaction schemes, have emerged as a versatile tool.
They are built on the idea to discretise in time a reduced dynamics, further initially simplifying to the extreme the modelling of the environment, to be seen as a collection of identical and independent units, later restoring in a controlled way a more realistic modelling, and possibly a continuous time limit.
This Perspective aims to be a primer for the interested reader to show how CMs provide a useful framework for modelling open system dynamics in a controllable manner, allowing for the introduction of simple mechanisms for memory effects and also providing an elegant means for exploring the thermodynamics of quantum systems. While we will provide the basic ingredients of CMs, in what follows we will focus on what we feel are some particularly insightful applications and refer e.g. to Refs.~\cite{Ziman2010,EspositoPRX,francescoQMQM} and the forthcoming comprehensive review~\cite{francescoReview} for more complete technical details.

\section{Collision Models}
At the price of some arbitrariness, we can set the birth of CMs as they are used in recent literature to a paper published by Jayaseetha Rau in 1963~\cite{Rau1963}. In this work she suggested a strategy for the treatment of the \textit{stochastic dynamics of quantum-mechanical systems}, i.e. an open quantum system dynamics in current terminology. Be it because the field of open quantum systems was still to take shape or be it because simple tools for numerical implementations were not as accessible as now, this seminal contribution was forgotten for decades, becoming appreciated only very recently.

CMs rose to popularity at the beginning of this century, starting with Refs.~\cite{ScaraniPRL, ScaraniPRA}. It is interesting to note that these contributions, while insisting on the same formalism as Rau, already brought in two important novel ingredients. On the one hand, they aimed to create a bridge between an open quantum system viewpoint and a quantum information one, where a CM picture naturally appears considering the repeated actions of a channel. In particular, attention was drawn to the role of quantum correlations, such as entanglement, which despite the concept tracing back to Schr\"odinger, still had yet to enter the stage when Ref.~\cite{Rau1963} appeared, let alone more general quantum information concepts. On the other hand, the setting was already a thermodynamic one, devoted to the study of a thermalising machine.

Other applications in open quantum systems soon appeared~\cite{ZimanPRA, Ziman2010} and were followed by a rapid growth of works exploiting a CM approach. A line of research that soon developed was
the use of CMs as a starting point for the derivation of master equations~\cite{PalmaPRL,Vacchini2013a,VacchiniPRL,Vacchini2014a,LorenzoPRA,ZimanPRA2017,LorenzoOSID,Filippov2020a}. The basic aim was to overcome the limitations intrinsic in the hyphothesis leading to master equations in Lindblad form, such as the separation of time scales, allowing to consider also situations
beyond weak  coupling and high temperature so as to include memory effects. As a result different new structures of master equations both in time-local and integro-differential form describing trace preserving and completely positive dynamics which are not of semigroup form have been devised.
A related use of CMs is as paradigm for the description of quantum transport~\cite{SabrinaArXiv, ShaoArXiv, LevyEPL}, in which the use of standard Lindblad master equations has also shown important shortcomings~\cite{Wichterich2007a}.
To name a few other directions we recall the use of CMs in the description of random interactions~\cite{Attal, PascazioPRL, GennaroEPL,GennaroPRA,CampbellPRA2009, Pellegrini}, in modelling quantum synchronisation~\cite{KarpatPRA}, information scrambling~\cite{JinPRAScramble}, thermometry~\cite{LandiThermometry}, quantum steering~\cite{StrunzSteering},
entanglement generation~\cite{CakmakGME}, stroboscopic implementation of and  non-Markovian effects on heat engines/refrigerators~\cite{PezzuttoEngine, LandiArXiv2, HuberCooling, AbahJPComm}, entropy production~\cite{LandiPRL2019}, classical objectivity~\cite{CampbellPRA2019, SabrinaPRR}, quantum memories~\cite{DeffnerPRA} and thermalisation~\cite{Parrondo, OnatEntropy}. 

Two important disclaimers have to be kept in mind. Firstly, as should be immediately obvious, the idea of modelling a physical phenomenon in terms of interactions taking place sequentially is recurrent in the literature and can be found, albeit with different formalisations, in many contexts. As an outstanding example we recall the case of the micromaser~\cite{Cresser1992a, Englert2002b}, in which the interaction of a mode of the electromagnetic field confined in a high-quality cavity with a collection of atoms flying through the cavity naturally fits the CM viewpoint and also allows a well-defined time continuous limit~\cite{VacchiniPRL}. A second crucial issue is connecting a CM description, where a notion of time is captured by the discrete number of steps passed, to a proper continuous time limit. These links are important for establishing a clear-cut connection to real physical systems, together with a microscopic interpretation. In this regard, such a viewpoint has already been successfully developed for the memoryless case, leading to measurement interpretations of a Markovian open quantum system dynamics~\cite{Barchielli2009,Wiseman2010}. It remains an intriguing and fundamental open question whether a CM approach can help in devising a measurement interpretation also for non-Markovian open system dynamics. 

\subsection{Basic framework}
\begin{figure}[t]
\centering
\includegraphics[width=\columnwidth]{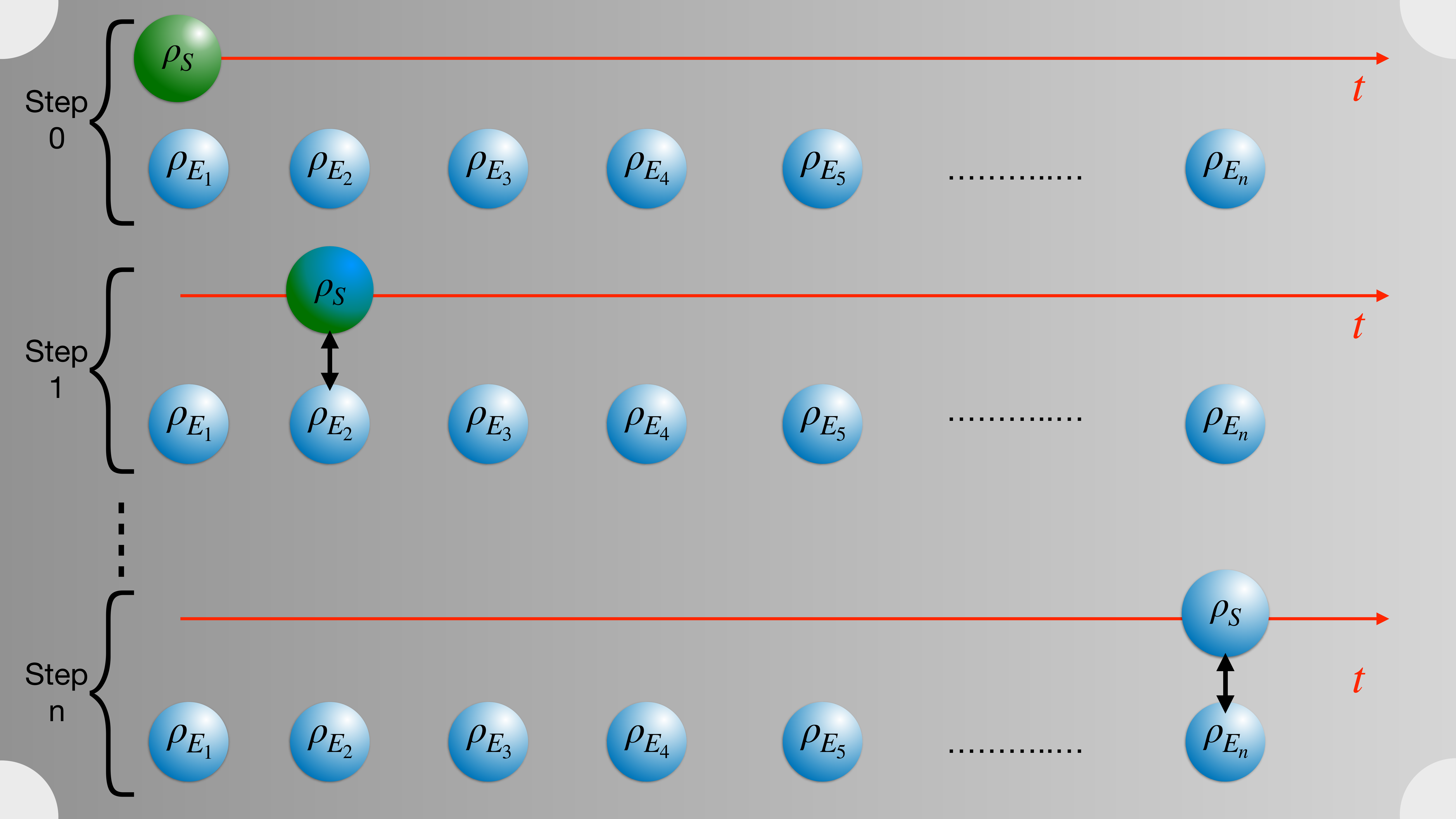}
\caption{CM schematics for the basic setting in which the system thermalises, with growing number of interactions, in a Markovian manner with the environment. The arrows denote time as described by number of collisions.}
\label{fig}
\end{figure}

In CMs one is interested in the dynamics of a quantum system, $S$, interacting with quantum environmental degrees of freedom, described making reference to two basic assumptions: \textit{i)} the environment is made up of a collection of identical auxiliary units, $E_i$, (often called ancillas); \textit{ii)} the system interacts sequentially and in a unitary fashion with distinct units. In the simplest setting one assumes all units to be initially uncorrelated, so that the total initial state reads
\begin{equation}
  \label{eq:1}
  \rho_{SE}(0)= \rho_{S}(0)\otimes\rho_{E_1}\otimes \rho_{E_2}\otimes\ldots \,,
\end{equation}
where all $\rho_{E_i}$ describe the same state,
and the interaction between system and $k$-th unit is described by the map $\mathcal{U}_{SE_{k}} $, corresponding to a fixed unitary evolution of duration $\tau$, the \textit{collision time}. After $n$ steps the reduced state of the system simply reads
\begin{equation}
  \label{eq:2}
  \rho_{S}(n \tau)= {\rm Tr}_{E_{n} \ldots E_{1}} \mathcal{U}_{SE_{n}}  \ldots \mathcal{U}_{SE_{1}}  \rho_{SE}(0) \equiv\Phi^n \rho_{S}(0) \,,
\end{equation}
where we have introduced the completely positive trace preserving map $\Phi$, $\Phi \sigma_S\equiv {\rm Tr}_{E} \mathcal{U}_{SE} \sigma_{S}\otimes\rho_{E}$, whose powers determine the system evolution. In physical terms, the system first interacts, for a fixed time and with arbitrary strength, with an environmental unit. It then moves on to undergo another collision with a distinct unit, uncorrelated from the previous one, thus conveying the idea of a memoryless bath, in which interaction events take place independently. This simple picture allows, for example, to account for thermalisation as described by a Lindblad master equation~\cite{ZimanPRA, francescoQMQM}, as visualised in Fig.~\ref{fig}.

A first advantage in the description of a dynamics in terms of CMs is their easy and natural numerical implementation, relying on the repeated application of a fixed transformation. Most importantly, additional freedom is available to modify, in a controlled way, features of the environment and of the interaction. Indeed, one can introduce initial correlations between system and environment, as well as correlations between environmental units. These correlations, that can be of both classical and quantum nature, play a crucial role in affecting the reduced system dynamics, and their occurrence is also related to the presence of memory effects, as we shall discuss in the next section. One can further consider different layers of environmental units as well as structured systems or a mediated interaction between system and units. All these different features allow for a rich modelling of the dynamics. In particular, in a CM one can partly or fully erase the correlations established between $S$ and $k$-th unit $E_k$ in between collisions, so as to make possible the assessment of the role of these correlations in influencing and characterising the reduced dynamics.  Two other important ingredients entering the description of the CM are the introduction of a random distribution of the elementary interaction events in time, as well as an intermediate time evolution or rearrangement of the environmental units in between subsequent collisions. Note that these collisions are often conveniently described in terms of a SWAP or CNOT operation, naturally arising in quantum information processing.

In view of this brief presentation, it appears that CMs prove to be useful in exploring the role of quantum coherences and quantum correlations in their interplay with environmental features in affecting properties of the reduced system dynamics. In this spirit we will now briefly introduce two research topics in which CMs have already been usefully exploited: quantum non-Markovianity and quantum thermodynamics.
\section{Non-Markovian dynamics}
A full-fledged treatment of open system dynamics besides the standard description of dissipation and decoherence effects must involve the analysis of memory effects. The characterisation of non-Markovianity in quantum systems is very challenging, since the classical probability treatment cannot be applied due to the roles of measurement and coherences. New strategies have therefore been devised~\cite{Rivas2014a}, and we will make reference to an approach based on the information exchange between system and environment~\cite{Breuer2016a}. The key quantity in this approach is a quantifier of the distinguishability between states $\mathcal{D}(\varrho,\sigma)$, contractive under the action of a completely positive trace preserving map and with the metric property
\begin{align}\label{shrinkTr-metric}
  \mathcal{D}(\Phi[\varrho], \Phi[\sigma])&\leqslant \mathcal{D}(\varrho,\sigma) \\
  \mathcal{D}(\varrho,\sigma)&\leqslant \mathcal{D}(\varrho,\tau)+\mathcal{D}(\tau,\sigma) \,,
\end{align}
such as the trace distance~\cite{Breuer2009b}
\begin{equation}
  \label{eq:4}
  \mathsf{D}(\varrho,\sigma)= 1/2\, \mathrm{Tr}\,|\varrho-\sigma| \,,
\end{equation}
or the square root of the Jensen-Shannon divergence
\begin{equation}
  \label{eq:3}
 \mathsf{J}(\varrho,\sigma) =\frac{1}{\sqrt{2}} \left[ \mathsf{S} \left(\varrho,\frac{\varrho+\sigma}{2}\right)+ \mathsf{S} \left(\sigma, \frac{\varrho+\sigma}{2}\right) \right]^{1/2} \,,
\end{equation}
with $\mathsf{S}(\cdot)$ the quantum relative entropy, as recently suggested~\cite{Megier2021a}.
The basic statement is that the distinguishability quantifier $\mathcal{D}(\varrho_{S}(t),\sigma_{S}(t))$,
with $\varrho_{S}(t)$ and $\sigma_{S}(t)$ states of the reduced system at time $t$ corresponding to the two distinct initial conditions $\varrho_{S}(0)$ and $\sigma_{S}(0)$,
provides the locally accessible information, doomed to decrease for a Markovian dynamics, in which the system looses information to the environment. In the presence of memory effects this distinguishability can undergo revivals, due to information coming back to the system and allowing to better tell the difference between two evolved system states. These revivals in time obey the following important upper bound \cite{Breuer2016a,Amato2018a}
\begin{align}\label{boundTr}
\!\!\!\!
  \mathcal{D}(\varrho_{S}(t),\sigma_{S}(t))-\mathcal{D}(\varrho_{S}(s),\sigma_{S}(s))\leqslant
  \mathcal{D}(\varrho_{E}(s),\sigma_{E}(s))+
  \nonumber
 \\ 
\mathcal{D}(\varrho_{SE}(s),\varrho_{S}(s)\otimes \varrho_{E}(s))+ \mathcal{D}(\sigma_{SE}(s),\sigma_{S}(s)\otimes \sigma_{E}(s)) \,,
\end{align}
where $t\geqslant s$ and $S,E$ denote system and environment respectively, telling us that memory effects are linked either to the establishment of system-environment correlations (recall that the initial state is assumed to be factorised), or to changes in the state of the environment.

Starting from~\cite{PalmaPRL}, CMs have proven a useful arena for the study of non-Markovianity in quantum system dynamics, in that they allow to engineer the details of the interaction, so as to be able to trace back the physical mechanisms leading to revivals in trace distance or other distinguishability quantifiers. The reference setting is always given by the paradigmatic Markovian modelling identified by Eqs.~\eqref{eq:1} and \eqref{eq:2} and depicted in Fig.~\ref{fig}. Memory effects can then be naturally introduced by removing independence between subsequent collisions.  A simple but effective strategy in this respect is the inclusion of interactions between neighbouring environmental units, which can be described by a stochastic SWAP operation, to be implemented either as a coherent or as an incoherent transformation taking place with a given probability~\cite{FrancescoPRA2013,FrancescoPhysScr,RuariPRA}, thus allowing to control its effect. The non-Markovianity introduced in such a way can be observed by studying as a function of time a suitable measure, e.g. the trace distance Eq.~\eqref{eq:4}, between initially distinct system states, memory effects being associated to revivals of this quantity. All the more, in these kinds of CMs one can also investigate how removing, partially or fully, the correlations established by the interaction between the system and a single environmental unit gradually washes out memory effects \cite{RuariPRA}. Collisions can also take place with many environmental units at once, thus allowing to model them as composite systems~\cite{JinNJP}, as well as more than one layer of environmental units, so as to describe a structured environment~\cite{LorenzoOSID}. This approach allows to better study the role of correlations and their partial erasure. In particular it allows to introduce additional figures of merit for non-Markovianity, such as the memory depth~\cite{CakmakPRA,CampbellPRA2018,PLAnonM,HuberCooling}, which quantifies the amount of environmental degrees of freedom that should be kept track of in order to obtain a Markovian description for the \textit{dressed system}. In such a way one can also clarify that it is not the presence of correlations per se which warrants non-Markovianity, but rather their influence on the reduced state, indeed different correlations and environmental changes can lead to the very same reduced dynamics \cite{Smirne2021a}. 
 
Clearly memory effects can be simulated in a variety of manners. Collisions taking place between environmental units before the system interacts with the subsequent unit can allow to accurately reproduce the dynamics of time-continuous non-Markovian systems, which typically corresponds to solutions of memory kernel master equations~\cite{VacchiniPRL,LandiArXiv}. This further allows to connect with repeated measurement models and quantum trajectories~\cite{PellegriniJPA , Vacchini2014a ,StrunzPRA, WhelanPRA}. In a similar framework one can also naturally introduce a notion of memory as an intermediate unit, possibly of a different dimension, that mediates the interaction between system and reservoir~\cite{SaroPRA}. Another natural strategy is the introduction of correlations in the initial environmental state~\cite{ZimanPRA2017, NadjaPRA, NadjaPRA2}. This medley of routes for introducing non-Markovianity in a controlled manner, having the freedom to change features of the environment or influence the established correlations, strengthens the connection between non-Markovianity and its physical interpretation in terms of the three contributions that upper bound the distinguishability revivals in Eq.~\eqref{boundTr}. These quantities act as precursors of non-Markovianity~\cite{CampbellNJP2019}, and allow for its quantitative study.

\section{Quantum thermodynamics}
Another area which has significantly benefited from both the simplicity and versatility of CMs is quantum thermodynamics. Consider the first law
\begin{equation}
\label{firstlaw}
\Delta E = \Delta W + \Delta Q
\end{equation} 
where $\Delta E$ is the total change in energy, $\Delta W$ is the work which is associated to changes in the total Hamiltonian and $\Delta Q$ accounts for the heat. For a single system weakly coupled to a thermal bath, these quantities can be readily determined solely based on the system degrees of freedom and a fully consistent thermodynamic framework can be established~\cite{Alicki}. However, in more general settings, ensuring the correct energetic and entropic accounting is crucial to avoid apparent violations of the thermodynamic laws. An exemplary case is when two mutually interacting systems of interest are coupled to their own respective thermal baths, with the baths at different temperatures. As shown in Ref.~\cite{LevyEPL}, if the dynamics is described by local Lindblad master equations, apparent violations of the second law occur. Despite describing a perfectly physical dynamics, the local description nevertheless fails to properly account for the energetic and entropic fluxes in play, and rather a global approach must be employed.

Remarkably, the local approach can be faithfully reconciled with the laws of thermodynamics by employing a CM description~\cite{GabrieleNJP}. Assuming the system is described by the total Hamiltonian $H_S \!=\! \sum_{i=1}^2 H_{S_i} + H_I$, where $H_I$ is the interaction between the subsystems, it can be shown that a CM with thermal environmental units and energy preserving system-unit interactions results, in the continuous time limit, in the local master equation description at first order in the system-reservoir interaction time~\cite{GabrieleNJP}. This is sufficient to ensure that local detailed balance is preserved, i.e. the energy extracted from the system by virtue of the collision is entirely transferred into the environmental unit. This condition can be succinctly expressed as $\left[ H_{S_i}+H_{E_i}, V_i \right]\!=\!0$, where $H_{E_i}$ is the environmental unit Hamiltonian and $V_i$ is the system-unit interaction. However, considering the total Hamiltonian, due to the interaction $H_I$ between the two systems, in general, global detail balance is lost, i.e. $\left[ H_{S}+H_{E}, V \right]\!=\!\left[ H_I, V \right] \!\neq\! 0$ and allows to reach non-equilibrium steady states. The discrepancy can then be fully accounted for by carefully assessing the energetics. In particular, while in the local master equation picture we are assuming all Hamiltonians are time-independent and therefore there is no work being performed, in the CM picture we clearly must account for the switching on and off of the system-environment interaction during each collision
\begin{equation}
\label{switchingwork}
\delta W = \int_{n\tau}^{(n+1) \tau} \left\langle \partial_t H_{tot} \right\rangle dt.
\end{equation} 
This ``switching" work is inaccessible in the local master equation description and therefore, despite being a convenient description for other dynamical features, the local approach misses key ingredients in describing the thermodynamics. However, when Eq.~\eqref{switchingwork} is properly accounted for one recovers the correct energy balance according to the first law, Eq.~\eqref{firstlaw}~\cite{GabrieleNJP, EspositoPRX, BarraSciRep}. Thus, when the switching work is identically zero the local master equation approach accurately captures also the thermodynamics. A similar picture applies when considering boundary driven many-body systems, where again the local approach generally fails to accurately describe the correct thermodynamics~\cite{BarraSciRep, PereiraPRE, PereiraPRE2}. 

Beyond providing an insightful route to resolving apparent inconsistencies in modelling the dynamics of open quantum systems, CMs provide a convenient and unified framework for the thermodynamic description of more general settings. As already noted, a single system of interest coupled to a large thermal bath will simply thermalise. However, if the system is also coupled to a stream of environmental units, the system is driven out of equilibrium and therefore, this kind of coupling, well modeled in a CM framework, can equally be viewed as either a second environment to which the system is coupled or a resource of non-equilibrium free energy. In order to correctly account for the energetic exchanges at play one must again carefully account for the switching work associated with the turning on and off of the system-environment interactions. By having access to some environmental degrees of freedom, a sharper formulation of the second law for non-equilibrium systems can then be derived~\cite{EspositoPRX}
\begin{equation}
\label{secondlaw}
\Sigma_S \equiv \Delta S_S + \Delta S_{E_i} - \beta Q \geq I_{S:E_i} \geq 0,
\end{equation}
where $\Sigma_S$ is the entropy production during a collision, $\Delta S_{S(E_i)}$ is the change in von Neumann entropy for the system (environmental unit), $Q$ is the heat exchanged with the thermal bath at inverse temperature $\beta$, and $I_{S:E_i}$ is the correlation, measured via the mutual information, established between $S$ and $E_i$ due to the collision. It is interesting to note that, since an environmental unit only ever interacts with the system once before being discarded, these correlations are immaterial in dictating the ensuing dynamics of the system, however, they play a key role in the thermodynamics~\cite{CampbellPRA2018}.

That CMs allow for (at least partial) access to the environmental degrees of freedom endows them with a greater scope than other more phenomenological approaches in establishing a consistent thermodynamic description of an open system dynamics. For a system coupled to a CM environment it is clear that there are {\it a priori} no constraints on the strength or form that the system-environment interaction can take, in particular one is not constrained to consider the partial SWAP for instance. This in turn allows for a complete characterisation of the available steady states and the associated thermodynamics in regimes beyond the scope of other approaches~\cite{CampbellPLA, BarraPRE, SeahPRE, StrasbergPRL2019}. For thermal environmental units and the partial SWAP interaction, the system will reach equilibrium, or homogenise, with the bath. This is the only interaction for which the switching work, Eq.~\eqref{switchingwork}, is identically zero. For other interactions the system is generally driven to a non-equilibrium steady state maintained by non-zero steady state heat and work fluxes. Once again, by virtue of the microscopic description afforded by the CM, the source of these fluxes is traced back to the work cost associated in switching on and off the system environment interaction~\cite{CampbellPLA, SeahPRE}, and thus, the CM provides an elegant demonstration of the house-keeping heat and work~\cite{CampbellPLA}. A wide range of non-equilibrium steady states are accessible, including those with coherence in the energy eigenbasis or temperature inverted states, simply by tuning this interaction term. Both these classes of states are so-called non-passive or active states which have a non-zero amount of extractable work associated to them. As such they are remarkable considering they can be seen as the result of a system coupled to a Markovian thermal bath. It is worth stressing this point. At first glance, the availability of such steady states might appear to violate the second law by allowing for extractable work from a single thermal bath. However, the CM description provides the necessary insight to reassess the energetics at play and reconcile the availability of these states with the known laws of thermodynamics, again, simply by accounting for the external work invested due to the switching on and off of the interaction.

The availability of non-passive steady states, and the associated thermodynamic analysis, has interesting practical considerations. By engineering the system environment interaction, a quantum system functioning as a battery can be charged via a dissipative process that, crucially, does not waste energy~\cite{BarraBattery}. The trick is to find a form of interaction that leads to a temperature inverted steady state, i.e. a steady state which does not require non-equilibrium heat or work fluxes to maintain it. Consider for example a system (environmental unit) with Hamiltonian $H_{S(E_i)}\!=\!\omega \sigma_z^{S(E_i)}$ with their mutual interaction, $V$. Assuming that all the $E_i$'s are in thermal states then if $\left[ H_S + H_{E_i}, V \right]\!=\!0$ the system will simply reach equilibrium and the associated heat and work fluxes vanish. However, for $V\!\!=\!\!\sigma_x\otimes\sigma_x -\sigma_y \otimes \sigma_y$, we find instead that $\left[ H_S - H_{E_i}, V \right]\!=\!0$, which drives the system to a temperature inverted steady state that, unlike other non-equilibrium steady states, has vanishing work and heat fluxes. Therefore, once the battery system is charged no further energetic cost is required to maintain it~\cite{BarraBattery}. 

The fact that a consistent thermodynamic framework can be readily derived by invoking a CM perspective further elevates their role in exploring other fundamental questions. A particularly fruitful line of work has employed CMs to better understand the delicate interplay between energy and information as captured by Landauer's principle, which states that the erasure of information comes at an inescapable thermodynamic cost of heat dissipated into the environment (or more accurately the non-information bearing degrees of freedom). For a single system in thermal equilibrium with a bath and considering the CM as an information reservoir~\cite{EspositoPRX}, Landauer's principle can be formulated in terms of the energy required to change the entropy of an environmental unit. If the system is assumed to already be in the steady state, we obtain a Landauer-like expression directly from Eq.~\eqref{secondlaw} by setting $\Delta S_S\!=\!0$. The insight gained from CMs reveals the non-trivial role that the $S$-$E_i$ correlations play, fully consistent with the quantum sharpening of Landauer's principle~\cite{ReebWolf}. In fact, through a CM approach these correlations have been shown to be highly relevant in extending and understanding information erasure and energy exchange in multipartite systems~\cite{LorenzoPRLLandauer} and the effects of structured environments and non-Markovianity~\cite{PezzuttoLandauer, SaroLandauer, Landauer3}.

\section{Final comments and future perspectives}
In this Perspective we have highlighted how CMs have provided a convenient testbed for the theoretical analysis of different phenomena in which the system of interest is interacting with other environmental degrees of freedom.
A central question which deserves a critical assessment is the adherence of CM evolutions to the dynamics of physical systems. In the first instance it is clear that, apart from special experimental configurations, the representation of a physical process by a discrete time description can only be an approximation. The conditions on the existence of a well-defined continuous time limit should clarify the class of dynamics that can be properly described by a given CM. Nevertheless, the versatility of CMs make them a very useful playground to understand which are the links between specific features of the environment as well as the interaction and the resulting reduced dynamics. This advantage is being thoroughly exploited in the investigation of both non-Markovianity and quantum thermodynamics. With regards to the former, the controlled insertion or removal of correlations and modification of coherence and correlation properties of the environment have been put in relation to the onset of memory effects. The analysis of the link between physical features and non-Markovianity in engineered CMs allows for an easier assessment and opens the way for their exploitation in real physical systems. While for the latter, access to environmental degrees of freedom permits for the proper assessment of the energetic and entropic exchanges, allowing for a consistent thermodynamic description. Let us stress that while a precise and complete correspondence between a CM and a physical system might not be available, CMs can still provide a useful tool for the evaluation of specific observables and/or figures of merit.

Finally, we note that recent advances have shown that several experimental platforms are very well suited to the implementation of CMs~\cite{SabrinaNPJ, MataloniSciRep, MarlettoArXiv} and that CMs can be combined with other numerically advanced techniques to tackle complex open system problems~\cite{PReB,GabrieleArXiv}. We therefore believe the diversity of phenomena that can be understood through the guise of a CM ensures that they will prove invaluable in all aspects of quantum simulation.

\acknowledgments
We are grateful to the many colleagues with whom we have shared discussions regarding these topics, in particular, Francesco Ciccarello, Gabriele De Chiara, Bar{\i}{\c{s}} {\c{C}}akmak, Giacomo Guarnieri, \"{O}zg\"{u}r E. M\"{u}stecapl{\i}o\u{g}lu, G. Massimo Palma, Mauro Paternostro and Andrea Smirne.  
S.C. acknowledges the SFI Starting Investigator Research Grant ``SpeedDemon" No. 18/SIRG/5508, B. V. acknowledges the UniMi Transition Grant H2020.

\end{document}